# Effects of Incomplete Ionization on β-$Ga_2O_3$ Power Devices: Unintentional Donor with Energy 110 meV

Adam T. Neal[1,2,*], Shin Mou[1,*], Roberto Lopez[3], Jian V. Li[3], Darren B. Thomson[4], Kelson D. Chabak[4], and Gregg H. Jessen[4]

[1] Air Force Research Laboratory, Materials and Manufacturing Directorate, Wright Patterson AFB, OH
[2] Universal Technology Corporation, Dayton, OH
[3] Texas State University, Department of Physics, San Marco, TX
[4] Air Force Research Laboratory, Sensors Directorate, Wright Patterson AFB, OH
[*] Electronic Address: shin.mou.1@us.af.mil and adam.neal.2.ctr@us.af.mil

Abstract:

Understanding the origin of unintentional doping in $Ga_2O_3$ is key to increasing breakdown voltages of $Ga_2O_3$ based power devices. Therefore, transport and capacitance spectroscopy studies have been performed to better understand the origin of unintentional doping in $Ga_2O_3$. Previously unobserved unintentional donors in commercially available ($\bar{2}$01) $Ga_2O_3$ substrates have been electrically characterized via temperature dependent Hall effect measurements up to 1000 K and found to have a donor energy of 110 meV. The existence of the unintentional donor is confirmed by temperature dependent admittance spectroscopy, with an activation energy of 131 meV determined via that technique, in agreement with Hall effect measurements. With the concentration of this donor determined to be in the mid to high $10^{16}$ cm$^{-3}$ range, elimination of this donor from the drift layer of $Ga_2O_3$ power electronics devices will be key to pushing the limits of device performance. Indeed, analytical assessment of the specific on-resistance ($R_{onsp}$) and breakdown voltage of Schottky diodes containing the 110 meV donor indicates that incomplete ionization increases $R_{onsp}$ and decreases breakdown voltage as compared to $Ga_2O_3$ Schottky diodes containing only the shallow donor. The reduced performance due to incomplete ionization occurs in addition to the usual tradeoff between $R_{onsp}$ and breakdown voltage. To achieve 10 kV operation in $Ga_2O_3$ Schottky diode devices, analysis indicates that the concentration of 110 meV donors must be reduced below $5\times10^{14}$ cm$^{-3}$ to limit the increase in $R_{onsp}$ to one percent.

While crystalline $Ga_2O_3$ has been known for many years, the recent availability of high quality crystalline substrates[1] and the demonstration of $Ga_2O_3$ MESFETs[2] and MOSFETs[3-6] have motivated interest in $Ga_2O_3$ for next generation ultra-wide bandgap power electronics applications. With a bandgap of 4.5 – 4.9 eV and estimated critical breakdown field of 8MV/cm, $Ga_2O_3$ possesses a Baliga figure of merit 10 times greater than SiC and four times greater than GaN.[2] Indeed, even at this early stage of development, electric fields of at least 3.8 MV/cm[4] and 5.1 MV/cm[7] have been demonstrated in lateral $Ga_2O_3$ FETs and vertical Schottky diodes, respectively, already surpassing bulk critical fields of GaN and SiC. In addition to these promising material properties, melt-growth methods for bulk $Ga_2O_3$ substrate growth[8-17] are expected to be more cost-effective than the sublimation techniques used for the growth of SiC substrates, lowering manufacturing costs for $Ga_2O_3$ based power electronics.

While the large breakdown electric field demonstrated in these early experimental studies is certainly promising, pushing the breakdown voltages of $Ga_2O_3$ based power electronics devices towards their predicted limits requires further work to understand and eliminate unintentional doping in the material. To illustrate this fact, the dependence of breakdown voltage of p-n single-sided junction and Schottky junction devices, normalized by material parameters, is plotted in Figure 1. The black line indicates the theoretical limit, assuming the full-depletion approximation and one dimensional electrostatics. By normalizing the breakdown voltages in this way, one can isolate the effect of doping on device breakdown voltage to allow comparisons of devices made from different materials. Comparing early $Ga_2O_3$ devices to those of more mature semiconductor materials gives insight into the future potential of $Ga_2O_3$ devices. The plotted experimental data are among the best reported for vertical devices of each material with large breakdown voltage and low doping of the device drift-layer. While edge effects prevent real devices of any material from matching the simplified theoretical limit exactly, the $N_d^{-1}$ trend is nevertheless observed considering the best performing devices of the different material systems. Comparing $Ga_2O_3$ to the more mature materials, it is clear that some improvement can be expected from the optimization of device geometries to mitigate edge effects, as shown by the blue arrow. However, much greater improvement in breakdown voltage is possible from reducing drift layer doping in $Ga_2O_3$ devices, as indicated by the orange arrow. For instance, if the drift layer doping of a $Ga_2O_3$ device can be reduced to about $10^{14}$ cm$^{-3}$ like the 21.7 kV SiC device indicated in Figure 1, then the $Ga_2O_3$ device will have a breakdown voltage over 100 kV as indicated by the right axis of Figure 1. Therefore, understanding and mitigating unintentional doping in $Ga_2O_3$ is key to increasing the maximum achievable breakdown voltages in $Ga_2O_3$ beyond the recently demonstrated 1 kV devices.[7]

With these motivations, we have undertaken transport and capacitance spectroscopy studies in order to better understand the origin of unintentional doping in $Ga_2O_3$. The transport properties of commercially available unintentionally doped ($\bar{2}$01) $Ga_2O_3$ substrates from Tamura Corporation grown via the EFG method[13,14] are characterized via temperature dependent Hall effect measurements. High temperature Hall effect measurements up to 1000 K reveal a previously unobserved unintentional donor with an energy level 110 meV below the conduction band edge in addition to previously observed shallow donors attributed to Si.[14,17] To confirm the existence of this unintentional donor, temperature dependent admittance spectroscopy measurements were performed on a $Ga_2O_3$ Schottky diode structure, with an activation energy of 131 meV observed via this technique. This activation energy agrees well with the donor energy of 110 meV determined via Hall effect measurements. Finally, using the information obtained from the temperature dependent Hall effect measurements, the effects of the 110 meV donor on the specific on-resistance ($R_{onsp}$) of $Ga_2O_3$ Schottky diodes are assessed via analytical calculations, indicating that incomplete ionization of the 110meV donor increases the $R_{onsp}$ and reduces breakdown voltage as compared to $Ga_2O_3$ devices with only the shallow donor, beyond the usual tradeoff between on-resistance and breakdown voltage.

**Results**

**Hall Effect Measurements.** Figure 2 shows the temperature dependent carrier density, normalized to room temperature, as measured by the Hall effect for two samples. The Hall scattering factor was assumed to be one when calculating the carrier density. The temperature dependent conductivity, mobility, and analysis of the relevant scattering mechanisms for these same samples can be found in the supplementary information. With the increased slope of the log-reciprocal plot from room temperature up to about 450 K, the temperature dependence of the carrier density must be determined by two donors with different energy levels. To estimate the donor energy levels, the data was fit with a model consisting of two donors with a compensating acceptor as shown in Equation (1):

$$N_c e^{\frac{E_f-E_c}{kT}} + N_a = \frac{N_{d1}}{1+2e^{\frac{E_f-E_{d1}}{kT}}} + \frac{N_{d2}}{1+2e^{\frac{E_f-E_{d2}}{kT}}} \tag{1}$$

where $N_c$ is the effective density of states in the conduction band, $N_{d1}$ and $N_{d2}$ the concentrations of the two donors, $N_a$ the concentration of compensating acceptors, $E_c$ the energy of the conduction band edge, $E_f$ the Fermi level, $E_{d1}$ and $E_{d2}$ the donor energies. An effective mass $m^* = 0.3m_o$[18-20] was used to estimate $N_c$ analytically,[21] while $N_{d1}$, $N_{d2}$, $E_{d1}$, $E_{d2}$, and $N_a$ are free parameters. The values of the free parameters are summarized in Table I, and the resulting fit is plotted as a black line in Figure 2. The model indicates that the increased slope from 300 K to 450 K is the result of a higher energy donor with energy 110 meV. Additionally, a shallow donor is also observed with energy 23 meV, previously identified as silicon from glow discharge mass spectrometry (GDMS) analysis of Tamura samples.[22] Carrier activation in previous studies of Ga$_2$O$_3$ was well described by a single shallow donor which they also attributed to unintentional silicon dopants;[14,17] however, the second higher energy donor has not been observed previously.

**Admittance Spectroscopy Measurements.** Admittance spectroscopy measurements confirm the existence of the previously unobserved donor. Figure 3 shows capacitance of a Ga$_2$O$_3$ Schottky diode as a function of frequency for several temperatures. The $C$ vs. $\omega$ spectrum exhibits a step transition separating two plateaus, $C_d$ at $\omega < \omega_p$ and $C_g$ at $\omega > \omega_p$, which indicates the presence of a trap state. The transition frequency, $\omega_p$, corresponds to the carrier emission rate from the trap level. To determine the activation energy of the trap state, $\omega_p$ was obtained from the negative peak in the differential capacitance ($\omega dC/d\omega$ vs $\omega$) spectrum to construct an Arrhenius plot of $\ln(\omega_p/T^2)$ vs $1/(k_BT)$, shown in Figure 3 inset. From the slope of the plot, the activation energy of the trap is determined to be $E_a = 131\pm5$ meV. Furthermore, the trap density, calculated from the height of the capacitance step,[23] is $N_t = 4.4\times10^{16}$ cm$^{-3}$. A relative dielectric constant of 10 for Ga$_2$O$_3$ was used for this calculation.[24,25] These data are well matched to the previously unobserved donor identified via Hall effect measurements, which indicated a donor energy of 110 meV and donor density of $7.5\times10^{16}$ cm$^{-3}$ as previously discussed. The small differences between the Hall effect measurements and admittance spectroscopy measurements are not surprising and are consistent with sample to sample variation within the two-inch Ga$_2$O$_3$ wafer from which the three samples were cut. Therefore, the trap state observed via admittance spectroscopy confirms the existence of the previously unobserved unintentional donor.

**Possible Origins of the Donor.** While the electrical characterization performed here cannot determine the chemical or crystallographic origin of the 110 meV donor, we nevertheless consider a few hypotheses. It is natural to ask if a native defect is responsible for the observed donor. DFT calculations rule out simple oxygen and gallium vacancies as they are determined to be deep donors[26] and acceptors,[27,28] respectively. However, antisites and interstitials could be responsible. Extrinsic impurities, of course, could also be responsible for the donor, and glow discharge mass spectrometry (GDMS) analysis of Tamura substrates indicates the presence of several impurities whose dopant properties in $Ga_2O_3$ are unknown.[22] Last, silicon on the octahedrally coordinated Ga(II) site of $Ga_2O_3$ could be responsible for the 110 meV donor. DFT calculations indicate that Si prefers the tetrahedrally coordinated Ga(I) site in which it is a shallow donor;[26] however, STM analysis of Si donors near the (100) surface of $Ga_2O_3$ indicates that Si occupies both Ga(I) and Ga(II) lattice sites.[29] This fact suggests that Si on the Ga(II) site should also be considered as a possible origin of the 110 meV donor.

**Discussion**

These unintentional donors have implications for the performance of $Ga_2O_3$ based power devices; namely they reduce the maximum achievable breakdown voltage in a $Ga_2O_3$ power device. Because the doping of the drift-layer cannot be reduced below the level of unintentional doping, the density of unintentional donors sets the lower limit on electric field at the metal-semiconductor interface and the upper limit on the depletion width of the $Ga_2O_3$ drift-layer for a given applied voltage. Therefore, a limit is set on the tradeoff which allows higher breakdown voltage to be achieved at the expense of larger on-resistance. To mitigate the presence of previously identified silicon shallow donors in the substrate, several methods including LPCVD,[30] MOCVD,[31,32] HVPE,[33,34] and MBE[2,3,35,36] have been used to grow homoepitaxial

Ga$_2$O$_3$ layers on Ga$_2$O$_3$ substrates, achieving lower carrier densities by minimizing unintentional doping. However, the carrier concentration due to unintentional doping still remaining in epitaxial layers is typically 10$^{15}$ to 10$^{17}$ cm$^{-3}$.[37] This concentration is similar to the additional carrier concentration due to the 110 meV donor observed in these bulk samples considering incomplete ionization. For example, at a 110 meV donor concentration of 7.5×10$^{16}$ cm$^{-3}$ as found in this study, with no other donors or acceptors, 3.4×10$^{16}$ cm$^{-3}$ free electrons are thermally activated into the conduction band at 300 K for an ionization efficiency of 46%. These facts suggest that the 110 meV donor may also play a role in the unintentional doping of epitaxially grown Ga$_2$O$_3$, warranting further study but beyond the scope of this work. The importance of the high temperature Hall effect measurements and admittance spectroscopy measurements should be emphasized. As our calculation indicates, the higher activation energy of the 110 meV donors means that their ionization efficiency $N_d^+/N_d$ is significantly less than 100% at room temperature. Therefore, room temperature and low temperature Hall measurement will not identify the full donor concentration, $N_d$, which determines the maximum breakdown voltage. Underestimating the concentration of donors could lead to an overestimate of the breakdown voltage without high temperature Hall effect measurements or admittance spectroscopy measurements like those performed in this study.

While unintentional shallow donors and the 110 meV donor both set a limit on the tradeoff between on-resistance and breakdown voltage, the 110 meV donor poses an additional challenge to maximizing breakdown voltage and minimizing on-resistance due to its four times larger donor energy as compared to the shallow donor. Because of the higher donor energy, incomplete ionization of the 110 meV donor is more severe than for the shallow donor, degrading the specific on-resistance vs. breakdown voltage characteristic for Ga$_2$O$_3$ based devices. The 110 meV donor is shallow enough in energy that it becomes fully ionized when the Schottky diode is reverse biased, reducing the depletion width, increasing the electric field at the metal-semiconductor interface, and reducing the breakdown voltage in the same way as the shallow donor. However, the 110 meV donor energy is large enough that incomplete ionization can be significant, meaning that only a fraction of the 110 meV donors are ionized when the Schottky diode is forward biased, reducing the carrier concentration available to conduct the on-current. To quantitatively examine those effects, specific on-resistance ($R_{onsp}$) versus breakdown voltage characteristics have been calculated for $Ga_2O_3$ Schottky diode devices including the incomplete ionization effect, with the results plotted in Figure 4. Details of the calculation can be found in the supplementary information. As Figure 4 shows, the on-resistance versus breakdown voltage characteristics are degraded where the maximum breakdown voltage decreases and the on-resistance increases as the concentration of the 110 meV donor increases. To determine the maximum acceptable concentration of 110 meV donors for a particular target breakdown voltage, the percent increase in $R_{onsp}$ is calculated for a Schottky diode containing both 110 meV donors and shallow donors versus a Schottky diode containing only shallow donors designed for the same breakdown voltage. The results of the calculation are plotted in Figure 5 with additional details available in the supplementary information. With Figure 5, we can estimate the maximum concentration of the 110 meV donors acceptable for 10 kV operation of $Ga_2O_3$ Schottky diode devices. To limit the increase in $R_{onsp}$ to one percent, the concentration of 110 meV donors must be less than $5\times10^{14}$ cm$^{-3}$. A similar analysis of the percent decrease in breakdown voltage due to 110 meV donors is also presented in the supplementary information.

In conclusion, high temperature Hall effect and admittance spectroscopy measurements have revealed a previously unobserved unintentional donor with energy 110 meV below the conduction band edge in commercially available unintentionally doped $Ga_2O_3$ substrates grown by the EFG method. The existence of these unintentional donors sets a limit on the maximum breakdown voltages that can be achieved in $Ga_2O_3$ devices and must be mitigated to achieve the full benefits of $Ga_2O_3$ based power electronics. Additionally, incomplete ionization of the 110 meV donor causes increased on-resistance and decreased breakdown voltage in diodes containing the donor, as compared to diodes containing only the shallow donor, beyond the usual tradeoff between on-resistance and breakdown voltage. To achieve 10 kV operation in $Ga_2O_3$ Schottky diode devices, analysis indicates that the concentration of 110 meV donors much be reduced below $5\times10^{14}$ $cm^{-3}$ to limit the increase in $R_{onsp}$ to one percent.

## Methods

**Sample Fabrication and Measurement**. Two van der Pauw test samples were diced from the same two inch wafer into 1 cm x 1 cm square pieces. Following dicing, samples were solvent cleaned and 50nm/1000nm Ti/Au contacts were sputtered on the sample corners. To improve contact resistance, the samples were annealed from room temperature up to 450°C with a 15 min ramp in a tube furnace with argon gas flow. A third sample from the same two inch wafer was prepared for admittance spectroscopy by depositing an indium tin oxide (ITO) transparent contact to form a $Ga_2O_3$ Schottky diode. Following device fabrication, temperature dependent van der Pauw and Hall effect measurements were carried out in two separate Hall effect measurement systems. An electromagnet with vacuum cryostat and closed-loop He refrigerator was used for measurements below room temperature, while an electromagnet with a quartz tube furnace with silicon carbide heater was used for measurements above room temperature. The samples were kept under nitrogen gas flow at atmospheric pressure during high temperature measurements. Admittance spectroscopy measurements were carried out under vacuum in a closed-loop He cryostat using a Keysight Agilent 4990A impedance analyzer.

**Data Availability.** The data that support the findings of this study are available on request from the corresponding author S.M.

## Acknowledgements

This material is based upon the work supported by the Air Force Office of Scientific Research under award number FA9550-17RXCOR438 and award number FA9550-15RYCOR163. The authors would like to thank Bill Mitchell and Said Elhamri for valuable discussions and Gerry Landis for experimental assistance.

## Author Contributions

S.M. initiated the project. A.T.N. and S.M. performed and analyzed Hall effect measurements. R.L. and J.V.L. performed and analyzed admittance spectroscopy measurements. A.T.N. and S.M. wrote the manuscript with input from J.V.L., D.B.T., K.D.C., and G.H.J. S.M. supervised the project.

Table I: Parameters for the carrier density vs. temperature model

| | | |
|---|---|---|
| $N_{d1}$ | (cm$^{-3}$) | $7.5 \times 10^{16}$ |
| $E_c - E_{d1}$ | (meV) | 110 |
| $N_{d2}$ | (cm$^{-3}$) | $1.4 \times 10^{17}$ |
| $E_c - E_{d2}$ | (meV) | 23 |
| $N_a$ | (cm$^{-3}$) | $1.0 \times 10^{16}$ |

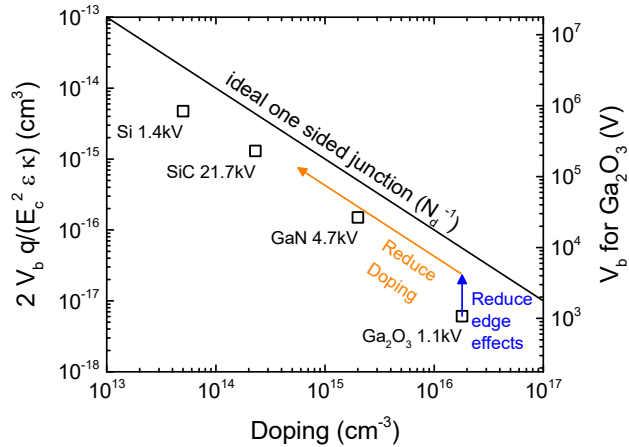

Figure 1: Normalized breakdown voltage (left axis) vs. doping concentration for single-sided junction power devices. The ideal relationship assuming the full-depletion approximation and 1D electrostatics is shown by the black line. The right axis indicates the projected absolute breakdown voltage for Ga$_2$O$_3$ devices for comparison. Square symbols indicate experimental measurements of devices with breakdown voltages among the highest reported for each material: Si[38], SiC[39], GaN[40], Ga$_2$O$_3$[7].

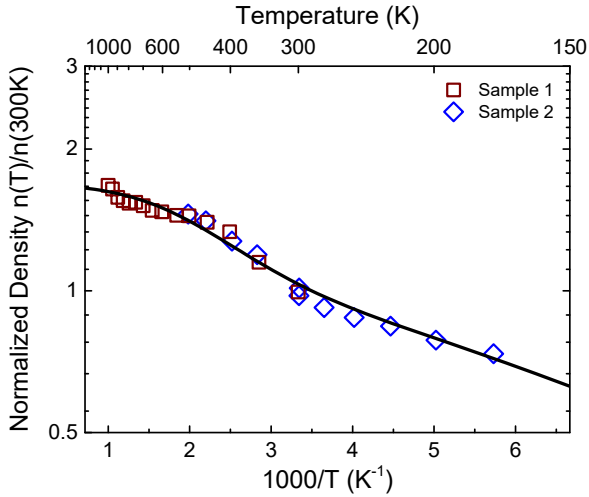

Figure 2: Hall carrier density (log scale) vs. 1000/T of $Ga_2O_3$ for two samples measured in the square geometry. The data are normalized to the Hall carrier density at 300 K. The symbols are the measured data and the black line a fit. Room temperature electron densities are $1.71 \times 10^{17}$ $cm^{-3}$ for Sample 1 (red square) and $1.21 \times 10^{17}$ $cm^{-3}$ for Sample 2 (blue diamond).

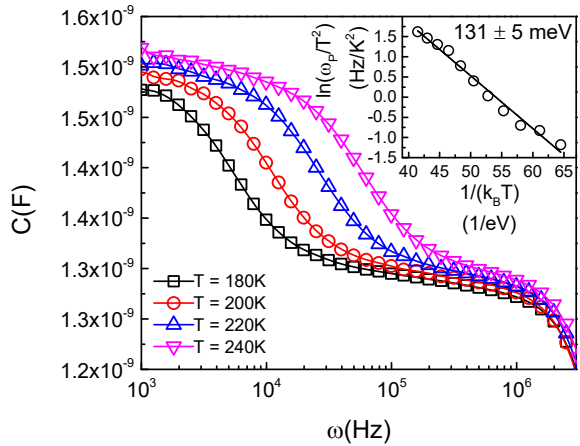

Figure 3: Frequency dependent capacitance (C-f) data at different temperatures (T) indicating the unintentional donor. Inset: Arrhenius plot [$\ln(\omega_p/T^2)$ vs $1/(k_BT)$] for the trap signature observed through admittance spectroscopy in which $\omega_p$ is the negative peak in the $\omega dC/d\omega$ vs $\omega$ spectrum. The activation energy is extracted from the slope (fitted line). This activation energy matches that determined for the unintentional donor via Hall effect measurement.

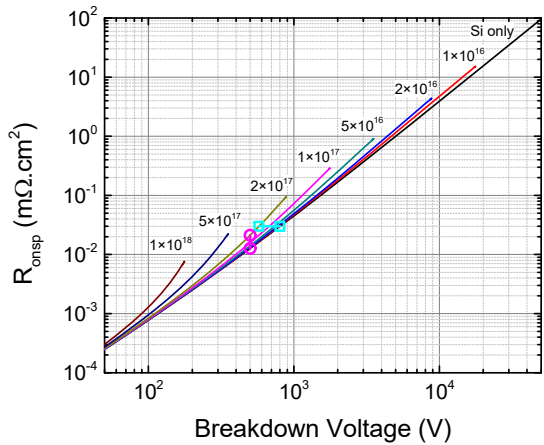

Figure 4: Analytical calculation of specific on-resistance ($R_{onsp}$) vs. breakdown voltage for $Ga_2O_3$ based Schottky junction devices with 110 meV donors including the effects of incomplete ionization. The label for each curve indicates the fixed concentration of 110 meV donors in units of $cm^{-3}$. The pink circles, cyan squares, and line segments illustrate the relationship to Figure 5 and Figure S3 as described in the supplementary information.

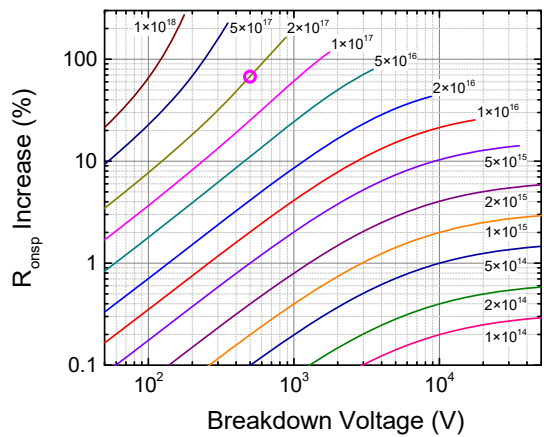

Figure 5: Percent increase in specific on-resistance ($R_{onsp}$) due to incomplete ionization as a function of breakdown voltage comparing $Ga_2O_3$ Schottky diodes with 110 meV donors to those without. The labels indicate the fixed concentration of 110 meV donors in $cm^{-3}$ for each curve. The pink circle illustrates the relationship to Figure 4 as described in the supplementary information.

Supplementary Information

# Effects of Incomplete Ionization on $\beta$-Ga$_2$O$_3$ Power Devices: Unintentional Donor with Energy 110 meV

Adam T. Neal[1, 2, *], Shin Mou[1, *], Roberto Lopez[3], Jian V. Li[3], Darren B. Thomson[4], Kelson D. Chabak[4], and Gregg H. Jessen[4]

[1] Air Force Research Laboratory, Materials and Manufacturing Directorate, Wright Patterson AFB, OH
[2] Universal Technology Corporation, Dayton, OH
[3] Texas State University, Department of Physics, San Marco, TX
[4] Air Force Research Laboratory, Sensors Directorate, Wright Patterson AFB, OH
[*] Electronic Address: shin.mou.1@us.af.mil and adam.neal.2.ctr@us.af.mil

**Temperature Dependent Conductivity and Hall Mobility**

The temperature dependence of the Hall mobility is well described by a combination of ionized impurity scattering and polar optical phonon scattering, with a measured mobility of about 430 cm$^2$/Vs at 140K, 160 cm$^2$/Vs at room temperature, and 20 cm$^2$/Vs at 1000K. Fitting of Hall mobility vs temperature yields a maximum room temperature mobility of 230 cm$^2$/Vs for Ga$_2$O$_3$, limited by polar optical phonon scattering, if impurity scattering is sufficiently reduced.

Figure S1 shows the conductivity for the same two samples for which Hall effect measurements are reported in the main text. Figure S2 show the Hall mobility, calculated from the conductivity in Figure S1 below and Hall carrier density from Figure 2 of the main text. The room temperature mobility is measured to be 160 cm$^2$/Vs, increasing to 430 cm$^2$/Vs at 140K. As temperature is increased, the mobility decreases to 20 cm$^2$/Vs. Because Ga$_2$O$_3$ is a compound semiconductor, polar optical phonon scattering is expected to dominate at high temperatures while impurity scattering is expected to dominate at low temperatures. Indeed, the temperature dependent mobility is well fit by the combination of these scattering mechanisms, in agreement with the recently published results of Ref 1. The momentum relaxation rates for screened ionized impurity scattering and polar optical phonon scattering, are as follows.[2] For screened ionized impurity scattering:

$$\frac{1}{\tau_{II}} = \frac{N_I q^4}{16\sqrt{2m^*}\pi\kappa_s^2\varepsilon_0^2}\left[\ln(1+\gamma^2) - \frac{\gamma^2}{1+\gamma^2}\right]E^{-3/2} \tag{S1}$$

$$\gamma^2 = \frac{8m^*EL_D^2}{\hbar^2} \tag{S2}$$

where $N_I$ is the number of ionized impurities and $L_D$ is the Debye screening length. For polar optical phonon scattering:

$$\frac{1}{\tau_{POP}} = \frac{q^2\omega_o\left(\frac{\kappa_S}{\kappa_\infty}-1\right)}{4\pi\kappa_S\varepsilon_0\hbar\sqrt{2[E/m^*]}}\left[N_o\sqrt{1+\frac{\hbar\omega_o}{E}} + (N_o+1)\sqrt{1-\frac{\hbar\omega_o}{E}} - \frac{\hbar\omega_o N_o}{E}\sinh^{-1}\left(\frac{E}{\hbar\omega_o}\right)^{1/2} \\ + \frac{\hbar\omega_o(N_o+1)}{E}\sinh^{-1}\left(\frac{E}{\hbar\omega_o}-1\right)^{1/2}\right] \tag{S3}$$

$$N_o = \frac{M}{e^{\hbar\omega_o/kT}-1} \tag{S4}$$

where $N_o$ is a semi-empirical distribution function for optical phonons with $M$ the effective number of optical phonon modes and $\hbar\omega_o$ the effective optical phonon energy. The total momentum relaxation rate is

$$\frac{1}{\tau_m} = \frac{1}{\tau_{II}} + \frac{1}{\tau_{POP}} \tag{S5}$$

From the solution of the Boltzmann transport equation in the relaxation time approximation for an energy dependent relaxation time, assuming Maxwell-Boltzmann carrier statistics, the mobility is[2]

$$\mu = \frac{q\langle\langle\tau_m\rangle\rangle}{m^*} \tag{S6}$$

$$\langle\langle\tau_m\rangle\rangle = \frac{\int_0^\infty E^{3/2}\tau_m(E)f(E)\,dE}{\int_0^\infty E^{3/2}f(E)\,dE} \tag{S7}$$

and $\langle\langle\tau_m\rangle\rangle$ is a weighted average of the momentum relaxation time over energy specific to transport calculations. The carrier distribution function, $f(E)$, it taken to be the equilibrium distribution function for the mobility calculation performed here. $M$, $\hbar\omega_o$, and $N_I$ are taken as free parameters to fit the experimental data, while others are estimated from the literature. Table S1 gives a summary of all parameters, and the mobility calculated using Equations S6 and S7 is shown as a solid black line in Figure S2. Individual components of the mobility associated with the two scattering mechanisms are also shown. The resulting fit indicates that ionized impurity scattering and polar optical phonon scattering appropriately describe the temperature dependent mobility of $Ga_2O_3$. Based on the fitting of polar optical phonon scattering in the $Ga_2O_3$ samples, it is estimated that the room temperature mobility could be increased from 160 cm$^2$/Vs to a maximum of 230 cm$^2$/Vs if the impurity concentration in the sample is further reduced. This fact is illustrated by the calculated polar optical phonon mobility plotted as the dashed line in Figure S2.

**Additional Details on the R$_{onsp}$ versus Breakdown Voltage Calculation**

To calculate the R$_{onsp}$ versus breakdown voltage characteristics of Figure 4 in the main text, ionized donor concentrations were calculated as a function of the total donor concentrations at room temperature using the charge neutrality equation

$$N_c e^{\frac{E_f - E_c}{kT}} + N_a = \frac{N_{d1}}{1 + 2e^{\frac{E_f - E_{d1}}{kT}}} + \frac{N_{d2}}{1 + 2e^{\frac{E_f - E_{d2}}{kT}}} \tag{S8}$$

with $N_a = 0$ and donor energies as specified in Table I of the main text. R$_{onsp}$ and breakdown voltage were calculated according to the following equations assuming the full-depletion approximation:

$$V_B = \frac{\kappa \, \varepsilon \, E_c^2}{2 \, q \, (N_{d1} + N_{d2})} \tag{S9}$$

$$R_{onsp} = \frac{\kappa \, \varepsilon \, E_c}{\mu \, q^2 \, (N_{d1} + N_{d2}) \, (N_{d1}^+ + N_{d2}^+)} \tag{S10}$$

The estimated breakdown field of 8 MV/cm for Ga$_2$O$_3$ was used for $E_c$. A room temperature polar optical phonon limited mobility of 230 cm$^2$/Vs was used for $\mu$, determined from our analysis of Ga$_2$O$_3$ scattering mechanisms above. Figure 4 of the main text shows R$_{onsp}$ versus breakdown voltage characteristics generated from Equations S9 and S10 by holding the 110 meV donor concentration $N_{d1}$ fixed while varying the shallow donor concentration $N_{d2}$.

**Additional Details on the Percent Increase in R$_{onsp}$ versus Breakdown Voltage Calculation due to 110 meV donors. Similar Calculation of Percent Decrease in Breakdown Voltage vs. R$_{onsp}$ due to 110 meV donors.**

To examine the change in performance of a Schottky diode when 110meV donors are present, the following procedure is adopted and used to generate Figure 5 of the main text and Figure S3 below. To characterize the change in performance, it is useful to compare the performance of a first Schottky diode device with both 110 meV donors and shallow donors to a second baseline Schottky diode device with only shallow donors. Furthermore, as device designers would typically adjust the donor concentration of the drift-layer to target a particular performance metric, breakdown voltage or on-resistance, it makes sense to compare two devices designed to have the same performance metric. Using this approach, we therefore compare a first Schottky diode with a particular concentration of 110 meV donors and a particular concentration of shallow donors to a second baseline Schottky diode with only shallow donors, where the concentration of shallow donors in the second baseline diode is adjusted to match either the breakdown voltage or on-resistance of the first diode. The results of those calculations are shown in Figure 5 of the main text and Figure S3 below. The x-axis of these figures indicates the performance metric, breakdown voltage or on-resistance, which is the same for the two diodes. The y-axis shows the percent change in the other performance metric comparing the two diodes. The calculations were performed for a series of 110 meV donor concentrations $N_{d1}$ ranging from $1\times10^{14}$ cm$^{-3}$ to $1\times10^{18}$ cm$^{-3}$ as labeled in the figures. Again, the shallow donor concentration $N_{d2}$ is adjusted for both diodes to match the particular breakdown voltage or R$_{onsp}$ shown on the x-axis. The symbols and line segments shown in Figures 4 and 5 of the main text and Figure S3 below graphically indicate the relationship between the three figures. The pink circles and vertical line segment in Figure 4 of the main text illustrate an example increase in $R_{onsp}$ corresponding to the pink circle plotted in Figure 5 of the main text. The cyan squares and horizontal line segment in Figure 4 of the main text illustrate an example decrease in breakdown voltage corresponding to the cyan square plotted in Figure S3 below. The symbols and line segments merely illustrate the relationship of Figure 4 to Figure 5 and Figure S3. The specific values chosen for the illustration are unimportant. With Figure 5 and Figure S3, we can estimate the maximum concentration of the 110 meV donors acceptable for 10 kV operation of $Ga_2O_3$ Schottky diode devices. Looking at 10 kV in Figure 5, the concentration of 110 meV donors must be less than $5\times10^{14}$ cm$^{-3}$ to limit the increase in $R_{onsp}$ to one percent. Similarly, considering the 10 kV $R_{onsp}$ of 4 mΩ.cm$^2$ in Figure S3, the concentration of 110 meV donors must be less than $1\times10^{15}$ cm$^{-3}$ to limit the decrease in breakdown voltage to one percent. Of course, this analysis assumes that the shallow donor density can also be sufficiently reduced to achieve 10 kV operation.

**Supplementary Information References**

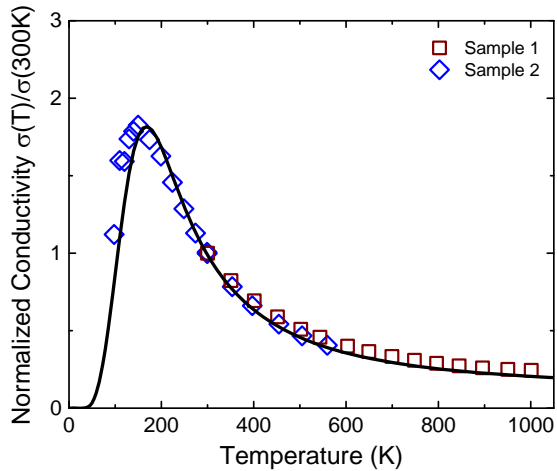

Figure S1: Conductivity vs. temperature of Ga$_2$O$_3$ for two samples measured by the van der Pauw method. The data are normalized to the conductivity at 300K. The symbols are the measured data and the black line a fit to the data. Room temperature conductivities are 3.75 S/cm for sample 1 (red square) and 3.11 S/cm for sample 2 (blue diamond).

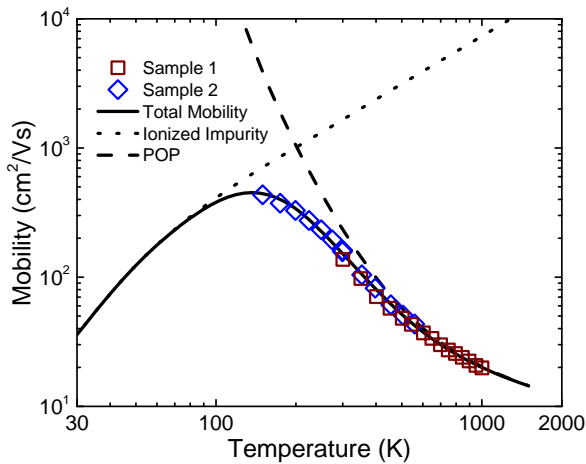

Figure S2: Log-Log plot of Hall mobility vs. temperature Ga$_2$O$_3$ for two samples determined from van der Pauw conductivity and Hall effect measurements. The symbols are the measured data and the black line a fit to the data. The two components of the mobility, screened ionized impurity scattering and polar optical phonon scattering, are plotted as dotted and dashed lines, respectively.

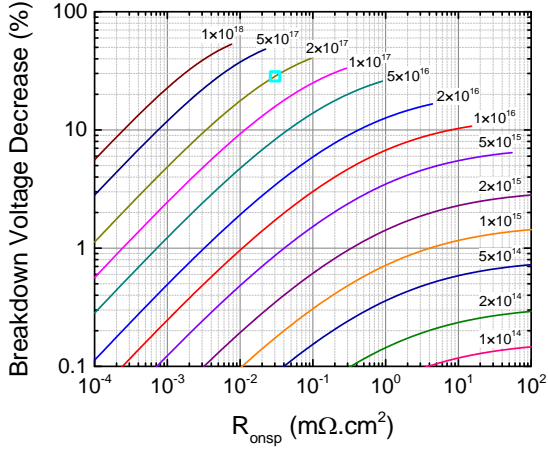

Figure S3: Percent decrease in breakdown voltage due to incomplete ionization as a function of specific on-resistance ($R_{onsp}$) comparing $Ga_2O_3$ based Schottky diode devices with both 110 meV donors and silicon donors to devices with only silicon donors. The labels indicate the fixed concentration of 110 meV donors in $cm^{-3}$ for each curve. Note that the percent decrease is calculated for Schottky diodes designed to have the same $R_{onsp}$, not silicon donor concentration. The cyan square is an example percent decrease in breakdown voltage which corresponds to the cyan squares and horizontal line segment in Figure 4 of the main text. The plotted symbol merely illustrates the relationship to Figure 4. The specific value chosen for the illustration is unimportant.

Table S1: Parameters for the mobility vs. temperature model

| | | |
|---|---|---|
| [a] $N_I$ | ($cm^{-3}$) | $1.8 \times 10^{17}$ |
| [a] $\hbar\omega_o$ | (meV) | 60. |
| [a] $M$ | | 1.7 |
| [b] $m^*/m_0$ | | 0.3 |
| [c] $\kappa_S$ | | 10 |
| [d] $\kappa_\infty$ | | 3.5 |

[a] free parameters, [b] Ref. 3,4,5 [c] Ref. 6,7
[d] Ref. 7,8,9